\documentclass[prd,twocolumn,showpacs]{revtex4}\begin{document}\preprint{}

\title{Stochastic inflation with quantum and thermal noise  }
\author{Z. Haba \\
Institute of Theoretical Physics, University of Wroclaw,\\ 50-204
Wroclaw, Plac Maxa Borna 9, Poland}
\email{zhab@ift.uni.wroc.pl}\date{\today}
\begin{abstract}
We add a thermal noise to Starobinsky equation of slow roll
stochastic inflation. We calculate the number of e-folds of the
stochastic system. The power spectrum and the spectral index are
evaluated from the fluctuations of the e-folds using an expansion
in the quantum and thermal noise terms.

\end{abstract}
 \maketitle

\section{Introduction}
\label{intro}

 The standard $\Lambda$CDM model describes the evolution of the universe in agreement with
 observations \cite{ade}. The fast expansion at the early stages
 of the evolution
 can be explained in terms of a scalar field (inflaton). A
 quantization of the scalar field and gravitational perturbations leads to fluctuations which can explain structure formation and the power spectrum of
  density
 fluctuations in the universe \cite{starfirst}\cite{mukhanov}\cite{starpl}\cite{hawking}\cite{guth}\cite{bar}. The model is introducing some new (dark) forms of
 matter and energy which are not interacting with the inflaton. If
 we assume that there are some interactions of the inflaton with
 the unknown forms of matter then the wave equation for the
 inflaton is transformed into a stochastic equation which
 in a flat expanding metric ( with the scale factor $a$ and $H=a^{-1}\partial_{t}a$) takes the form
 \begin{equation}
\partial_{t}^{2}\phi-a^{-2}\triangle\phi+(3H+\gamma^{2})\partial_{t}\phi+V^{\prime}(\phi)+\frac{3}{2}\gamma^{2}H\phi=\gamma a^{-\frac{3}{2}}\eta,
 \end{equation}
 where $\gamma^{2}$ is a friction related to the Gaussian noise
\begin{equation}
\langle \eta(t)\eta(s)\rangle =\delta(t-s)\end{equation} according
to the fluctuation-dissipation relation. Eq.(1) has been derived
in \cite{berera}(see also \cite{habacta}).  The friction
$\gamma^{2}$ is proportional to temperature. The noise $\eta$
comes from the thermal (Gibbs) distribution of the initial
positions  and velocities of the particles of the environment ( in
general, the environment may consist of any degrees of freedom
which are unobservable and averaged in a description of an
interaction with $\phi$). Eq.(1)is a basis of the warm inflation
\cite{warm}. In such a model the resonant reheating is unnecessary
as the temperature during inflation does not fall to zero owing to
the creation of radiation as a result of the decay of the
inflaton. The quantum fluctuations of the inflaton and
gravitational perturbations are usually described
\cite{mukhanov}\cite{bar} in a linear approximation. Starobinsky
\cite{starpl}\cite{starob} (see also \cite{vilenkin}) discovered
that  quantum fields at large time in an expanding universe behave
like a classical diffusion process. Then, the high momentum part
of the quantum field can be treated as an additional (quantum)
noise in the inflaton wave equation. Such a treatment of quantum
fluctuations during inflation goes beyond a linear approximation.
The quantum noise has been widely studied in
refs.\cite{linde}-\cite{vennini}. The Fokker-Planck equation for
the probability distribution of the inflaton has been explored in
detail. In principle, the Fokker-Planck equation contains all the
information about the probability distribution. In particular, the
power spectrum of fluctuations could be calculated  as discussed
in \cite{fin1}\cite{fin2} (see also Appendix B here). However, in
\cite{vennini} (following \cite{starob}, see also
\cite{ven1}\cite{ven2}) an alternative method has been  proposed
for a calculation of the power spectrum of the quantum noise based
on fluctuations of the e-folds. In this paper we extend the method
to the calculation of the power spectrum of the system which
contains both the quantum noise and the thermal noise. The plan of
this paper is the following. In sec.2 we discuss the stochastic
equation with the quantum and thermal noises in the slow roll
approximation.  In sec.3 following
refs.\cite{vennini}\cite{gikhman} we obtain general formulas for
the expectation values of e-folds and the fluctuations of e-folds
(spectral function). Then, in sec.4 we discuss approximations
leading to some explicit formulas for the spectral function and
the spectral index. In Appendices A and B we discuss fluctuations
in soluble models and the relation between Ito and Stratonovitch
stochastic equations.
\section{Slow roll stochastic equations}
We consider two sources of noise in eq.(1) the thermal noise
$\eta$ and the quantum noise $\eta_{S}$
\cite{starpl}\cite{starob}\cite{vilenkin}. The quantum noise comes
from the large momentum part (above the Hubble horizon) of the
scalar field. It can be considered as a part of the quantum
inflaton equation. The thermal noise results from an approximation
of the interaction with an environment by a Markov process. In
Einstein equations the environment could be represented as dark
matter or
 dark energy if we properly  choose the environmental interactions \cite{ha1}\cite{hss}\cite{sudarsky}.
 We consider eq.(1) in the
slow-roll approximation \begin{equation} (3H+\gamma^{2})\circ
d\phi = -V^{\prime}dt -\frac{3}{2}\gamma^{2}H\phi dt+\gamma
a^{-\frac{3}{2}}\circ dB+\frac{3}{2\pi}H^{\frac{5}{2}}\circ dW
\end{equation}In eq.(3) we write $\eta=\partial_{t}B$, $\eta_{S}=\partial_{t}W$ and assume that $W$ and $B$ are independent Gaussian
variables. We use the notation $\circ dW$ (after \cite{ikeda} )
for the Stratonvitch interpretation of the stochastic differential
and the conventional notation of the differential in the Ito
interpretation.  The difference between Ito and Stratonovitch
integrals consists in a different discrete time approximations of
the Riemann sums approximating the integral. The Stratonovitch
integral $\int f\circ dW$ treats time approximation of $f$ and $W$
(in the Riemann sum) in a symmetric way (so called middle point
approximation) whereas in the Ito integral the time in $dW$ is
later than in $f$. For an integral with a differentiable function
$W$ the various discrete approximations would lead to the same
result. However, $W$ is not differentiable. We discuss both
interpretations of the stochastic differential for an easy
comparison with literature on the subject. The Ito stochastic
differential equation can be expressed by the Stratonovitch
equation using the rule \cite{ikeda} $f\circ dW= fdW+
\frac{1}{2}dfdW$. So, both equations differ by a correction term.
The Stratonovitch form is convenient for calculations because it
preserves the standard rules of differentiation  (the Leibniz
rule)\cite{ikeda}. It must be checked in mathematical models which
form of the stochastic equation better describes physical
processes.

  In
order to simplify further discussion we assume that
$3H>>\gamma^{2}$ and $V^{\prime}>>\frac{3}{2}\gamma^{2}H\phi$.The
friction term $\gamma^{2}\partial_{t}\phi$ is usually related to
the decay of the inflaton into radiation \cite{rad}. The omission
of the $\gamma^{2}$ terms on the lhs of eq.(1) means a negligible
density of radiation (which can be true during inflation
\cite{fang}).
 Now, the stochastic
equation (3) reads
\begin{equation}
d\phi=-\frac{1}{3H}V^{\prime}dt+\frac{\gamma}{3}
a^{-\frac{3}{2}}H^{-1}\circ dB(t)
+\frac{1}{2\pi}H^{\frac{3}{2}}\circ dW(t).
\end{equation}The Starobinsky \cite{starob}\cite{vilenkin}
 slow-roll (quantum) system corresponds to the limit
$\gamma\rightarrow 0$ of eq.(4). In order to obtain an agreement
with the inhomogeneous inflaton and gravity perturbations (most
easily treated in the uniform curvature gauge
\cite{hwang}\cite{mar}) we must change the world time $t$ into the
e-folding time $\nu$ \cite{fin1}\cite{fin2}\cite{vennini}(usually
denoted by $N$; we change notation for typographical reasons)
describing the change of the scale factor
\begin{equation}
\nu=\int_{0}^{t}Hds=\ln(\frac{a}{a_{0}}).
\end{equation}
Now, the diffusion (small roll) system reads
\begin{equation}
 d\phi = -\frac{1}{3}V^{\prime}H^{-2}d\nu +\frac{\gamma}{3}
a^{-\frac{3}{2}}H^{-\frac{3}{2}}\circ dB(\nu)+\frac{1}{2\pi}H\circ
dW(\nu),
\end{equation}together with the differential form of the Friedman
equation ( taking a derivative in the Friedman equation can allow
to treat the environment   as a dark energy \cite{ha1})
\begin{equation}
d\ln(H)=-4\pi G(\partial_{\nu}\phi)^{2}d\nu.
\end{equation}
We can insert in eq.(4) either $a(\phi)$ as a function of $\phi$
or $a=a_{0}\exp(\nu)$ (in such a case we obtain a non-stationary
stochastic equation).

 In the slow-roll approximation we can derive from eq.(7) in the no-noise limit
\begin{equation}
H= \sqrt{\frac{8\pi G}{3}V},\end{equation} ( $V$ entering eq.(6)
can be determined by $H$ from the wave equation up to an arbitrary
constant). Then, from eqs.(4)-(5) in the no noise limit
\begin{equation} \ln (a)=-8\pi G\int
d\phi(V^{\prime})^{-1}V.
\end{equation} We could take in eq.(9) the noise into account by
means of perturbation methods (the relation between $a(\nu)$ and
$\phi(\nu)$ will still be discussed in Appendices A and B).

 The probability distribution  of the solution of eq.(4) (Stratonovitch
 interpretation)
 satisfies the Fokker-Planck
equation \cite{ikeda}\cite{gikhman} \cite{risken}
\begin{equation}\begin{array}{l}
\partial_{t}P=\partial_{\phi}\frac{\gamma^{2}}{18Ha^{\frac{3}{2}}}\partial_{\phi}\frac{1}{Ha^{\frac{3}{2}}}P
+\frac{1}{8\pi^{2}}\partial_{\phi}H^{\frac{3}{2}}\partial_{\phi}H^{\frac{3}{2}}P
\cr+\partial_{\phi}(3H)^{-1}V^{\prime}P. \end{array}\end{equation}
In the Ito interpretation of eq.(4)
\begin{equation}\begin{array}{l}
\partial_{t}P=\partial_{\phi}\partial_{\phi}\frac{\gamma^{2}}{18H^{2}a^{3}}P
+\frac{1}{8\pi^{2}}\partial_{\phi}\partial_{\phi}H^{3}P
\cr+\partial_{\phi}(3H)^{-1}V^{\prime}P .\end{array}\end{equation}
If in the e-folding time we treat $a$ as depending on $\phi$ (not
on $\nu$), then we obtain a stationary form of the Fokker-Planck
equation which for the Ito version is

\begin{equation}\begin{array}{l}
\partial_{\nu}P=\frac{\gamma^{2}}{18}\partial_{\phi}\partial_{\phi}\frac{1}{a^{3}H^{3}}P
 +\frac{1}{8\pi^{2}}\partial_{\phi}\partial_{\phi}H^{2} P
+\partial_{\phi}(3H^{2})^{-1}V^{\prime}P
\end{array}\end{equation}
and Stratonovitch version\begin{equation}\begin{array}{l}
\partial_{\nu}P=\frac{\gamma^{2}}{18}\partial_{\phi}\frac{1}{(Ha)^{\frac{3}{2}}}\partial_{\phi}\frac{1}{(Ha)^{\frac{3}{2}}}P
+\frac{1}{8\pi^{2}}\partial_{\phi}H\partial_{\phi}H P
\cr+\partial_{\phi}(3H^{2})^{-1}V^{\prime}P.
\end{array}\end{equation} If we  express $a$ by the $\nu$ time
then eq.(13) reads
\begin{equation}\begin{array}{l}
\partial_{\nu}P=\frac{\gamma^{2}}{18}\exp(-3\nu)\partial_{\phi}\frac{1}{H^{\frac{3}{2}}}\partial_{\phi}\frac{1}{H^{\frac{3}{2}}}P
+\frac{1}{8\pi^{2}}\partial_{\phi}H\partial_{\phi}H P
\cr+\partial_{\phi}(3H^{2})^{-1}V^{\prime}P
\end{array}\end{equation}
It can be seen from eqs.(6) and (14) that quantum and thermal
fluctuations are of the same order if $\gamma^{2}\simeq
e^{3\nu}H(\nu)^{5}$. The estimate of the dissipation strength
$\gamma$ is relevant for the estimate of the power spectrum and
the spectral index at the end of sec.4.

 Eq.(14) does not depend on the $a(\phi)$
approximation.
 $H(\phi)$  as a function of $\phi$ in eqs.(10)-(14) can be obtained from eq.(8). The dependence
of $a$ on $\phi$ in eqs.(10)-(13) in the slow-roll approximation
is determined by eq.(9). Let us consider simple examples. If
$V=g\phi^{n}$ (chaotic inflation \cite{linde2}) then
\begin{equation}
\frac{a(\phi)}{a_{0}}=\exp\Big(-4\pi Gn^{-1}\phi^{2}\Big).
\end{equation}
If $V=g\exp(\lambda\phi)$ then
\begin{equation}
\frac{a(\phi)}{a_{0}}=\exp\Big(-\frac{8\pi
G}{\lambda}\phi\Big),\end{equation}Note that if $\phi\rightarrow
+\infty$ then $a\rightarrow 0$, if $\phi\rightarrow -\infty$ then
$a\rightarrow \infty$ .

For a flat potential
\begin{equation}
V=\frac{\phi^{2}}{K+\phi^{2}}
\end{equation}
we have
\begin{equation}\begin{array}{l}
\frac{a(\phi)}{a_{0}}=\exp\Big(-2\pi G\phi^{2}-\frac{\pi
G}{K}\phi^{4}\Big).\end{array}\end{equation} For "natural
inflation" \cite{natural}
\begin{equation}
V=g(1-\cos\phi).
\end{equation}
Then,
\begin{equation}
a(\phi)=a_{0}\exp\Big(8\pi
G\ln\Big(2\cos^{2}(\frac{\phi}{2}\Big)\Big).
\end{equation}
 In the case of the double-well potential
\begin{equation}
V(\phi)=\frac{g}{4}(\phi^{2}-\frac{\mu^{2}}{g})^{2}
\end{equation}
\begin{displaymath}\begin{array}{l}
a=a_{0}\vert\phi\vert^{\frac{2\pi G\mu^{2}}{g}}\exp(-\pi
G\phi^{2}).
\end{array}\end{displaymath}
The noise corrected $a(\phi)$ relation could in principle be
derived from the solution of eq.(6) but in general this is
possible only on a perturbative level (see Appendix A).
\section{Expectation value of e-folds}We treat $\nu$ (in eq.(5)) as a
random time (because $a$ is random). Let us consider a
differential of a function of the stochastic process (6) in the
e-folding time in the Stratonovitch sense
\begin{equation}\begin{array}{l}
df=\partial_{\phi}f\circ d\phi\cr=\partial_{\phi}f\circ(
-\frac{1}{3H^{2}}V^{\prime}d\nu
+\frac{\gamma}{3(aH)^{\frac{3}{2}}}dB(\nu)+\frac{1}{2\pi}
HdW(\nu)) \cr =\partial_{\phi}f( -\frac{1}{3H^{2}}V^{\prime}d\nu
+\frac{\gamma}{3(aH)^{\frac{3}{2}}}dB(\nu)+\frac{1}{2\pi}
HdW(\nu)\cr
+\Big(\frac{1}{2}\frac{\gamma}{3(aH)^{\frac{3}{2}}}\partial_{\phi}\frac{\gamma}{3(aH)^{\frac{3}{2}}}\partial_{\phi}f
+\frac{1}{8\pi^{2}}\partial_{\phi}H\partial_{\phi}Hf\Big)d\nu.
\end{array}\end{equation}
For the Ito stochastic equation
\begin{equation}\begin{array}{l}
df=\partial_{\phi}f\circ d\phi\cr=\partial_{\phi}f\circ(
-\frac{1}{3H^{2}}V^{\prime}d\nu
+\frac{\gamma}{3(aH)^{\frac{3}{2}}}dB(\nu)+\frac{1}{2\pi}
HdW(\nu)) \cr =\partial_{\phi}f( -\frac{1}{3H^{2}}V^{\prime}d\nu
+\frac{\gamma}{3(aH)^{\frac{3}{2}}}dB(\nu)+\frac{1}{2\pi}
HdW(\nu))\cr
+\Big(\frac{1}{2}\Big(\frac{\gamma}{3(aH)^{\frac{3}{2}}}\Big)^{2}\partial_{\phi}\partial_{\phi}f
+\frac{1}{8\pi^{2}}H^{2}\partial_{\phi}\partial_{\phi}f\Big)d\nu.
\end{array}\end{equation} In the rest of this section we follow refs.\cite{vennini}\cite{gikhman}\cite{venninith}.
 In the Stratonovitch case we choose
a function $f$\begin{equation}\begin{array}{l}
-\partial_{\phi}f_{S}\frac{1}{3H^{2}}V^{\prime} +
\frac{1}{18}\frac{\gamma}{(aH)^{\frac{3}{2}}}\partial_{\phi}\frac{\gamma}{(aH)^{\frac{3}{2}}}\partial_{\phi}f_{S}
\cr+\frac{1}{8\pi^{2}}\partial_{\phi}H\partial_{\phi}Hf_{S}=-1
\end{array}\end{equation}and in the Ito case
\begin{equation}\begin{array}{l}
-\partial_{\phi}f_{I} \frac{1}{3H^{2}}V^{\prime}+
\frac{1}{18}\Big(\frac{\gamma}{(aH)^{\frac{3}{2}}}\Big)^{2}\partial_{\phi}\partial_{\phi}f_{I}
+\frac{1}{8\pi^{2}}H^{2}\partial_{\phi}\partial_{\phi}f_{I}=-1.
\end{array}\end{equation}Then, integrating $df$ between $\nu=0$ and $\nu$
corresponding to the values $\phi(0)=\phi_{in}$ and
$\phi(\nu)=\phi$ we obtain (the expectation value of the Ito
integral is equal to zero)
\begin{equation}
\langle \nu\rangle= \langle f(\phi_{in})\rangle -\langle
f(\phi)\rangle.\end{equation} Let $\partial_{\phi}f=u$ .
Eqs.(24)-(25) for $u$ are of the form
\begin{equation}
\partial_{\phi}u+Q(\phi)u=-r,
\end{equation}
where \begin{equation}
 Q=-\frac{1}{3H^{2}}V^{\prime}\Big(\frac{H^{2}}{8\pi^{2}}+\frac{\gamma^{2}}{18a^{3}H^{3}}\Big)^{-1}
 \end{equation}
 and (Ito interpretation)
 \begin{equation}
 r=\Big(\frac{H^{2}}{8\pi^{2}}+\frac{\gamma^{2}}{18a^{3}H^{3}}\Big)^{-1}.
\end{equation}
 The solution  of eq.(27) is \begin{equation}
u(\phi)=-\int_{\phi_{*}}^{\phi}d\psi
r(\psi)\exp\Big(-\int_{\psi}^{\phi}Q(X)dX\Big),
\end{equation}where $\phi_{*}$ is chosen to satisfy proper boundary conditions.

Then ( an analog of the formula derived by Starobinsky and Vennin
\cite{vennini})
\begin{equation}
f(\phi)=-\int_{\phi_{1}}^{\phi}d\phi^{\prime}\int_{\phi_{*}}^{\phi^{\prime}}d\psi
r(\psi)\exp\Big(-\int_{\psi}^{\phi^{\prime}}Q(X)dX\Big).
\end{equation}This solution satisfies  $f(\phi_{1})=0$ and
$\phi_{*}$ is chosen so that $f(\phi_{2})=0$. Then, according to
eq.(26)(setting $\phi=\phi_{2}$ to get $f(\phi_{2})=0$ ) we have
(an approximate formula for a mean value of e-folds appeared
already in \cite{starob})\begin{equation} \langle
\nu\rangle=-\int_{\phi_{1}}^{\phi_{in}}d\phi^{\prime}\int_{\phi_{*}}^{\phi^{\prime}}d\psi
r(\psi)\exp\Big(-\int_{\psi}^{\phi^{\prime}}Q(X)dX\Big).
\end{equation}$\nu$ is the umber of e-folds between
$\phi_{in}$ and $\phi_{end}=\phi_{1}$. We have to determine
$\phi_{*}$ from the condition $f(\phi_{2})=0$.

Another method is considered in \cite{gikhman}. There, the
solution of eq.(27) is written in the form
\begin{equation}\begin{array}{l}
u(\phi)=\exp\Big(-\int_{\phi_{*}}^{\phi}Q(X)dX)\Big)u(\phi_{*})\cr-\int_{\phi_{*}}^{\phi}d\psi
r(\psi)\exp\Big(-\int_{\psi}^{\phi}Q(X)dX\Big)\end{array}\end{equation}
Then, the boundary conditions are expressed by $u(\phi_{*})$ and
$\phi_{*}$. Integrating eq.(33)
\begin{equation}\begin{array}{l}
f(\phi)=\int_{\phi_{1}}^{\phi}d\phi^{\prime}\exp(-\int_{\phi_{*}}^{\phi^{\prime}}Q)u(\phi_{*})\cr-
\int_{\phi_{1}}^{\phi}d\phi^{\prime}\int_{\phi_{*}}^{\phi^{\prime}}d\psi
r(\psi)\exp(-\int_{\psi}^{\phi^{\prime}}Q).\end{array}
\end{equation}We demand
\begin{equation}\begin{array}{l}
0=f(\phi_{2})=\int_{\phi_{1}}^{\phi_{2}}d\phi^{\prime}\exp(-\int_{\phi_{*}}^{\phi^{\prime}}Q)u(\phi_{*})\cr-
\int_{\phi_{1}}^{\phi_{2}}d\phi^{\prime}\int_{\phi_{*}}^{\phi^{\prime}}d\psi
r(\psi)\exp(-\int_{\psi}^{\phi^{\prime}}Q)
\end{array}\end{equation}Solving for $u(\phi_{*})$ gives the formula for
$\nu$
\begin{equation}\begin{array}{l} \langle \nu\rangle=
\Big(\int_{\phi_{1}}^{\phi_{in}}d\phi^{\prime}\exp(-\int_{\phi_{*}}^{\phi^{\prime}}Q)
\cr\int_{\phi_{1}}^{\phi_{2}}d\phi^{\prime}\int_{\phi_{*}}^{\phi^{\prime}}d\psi
r(\psi)\exp(-\int_{\psi}^{\phi^{\prime}}Q)\cr-
\int_{\phi_{1}}^{\phi_{2}}d\phi^{\prime}\exp(-\int_{\phi_{*}}^{\phi^{\prime}}Q)
\int_{\phi_{1}}^{\phi_{in}}d\phi^{\prime}\int_{\phi_{*}}^{\phi^{\prime}}d\psi
r(\psi)\exp(-\int_{\psi}^{\phi^{\prime}}Q)\Big)\cr
\Big(\int_{\phi_{1}}^{\phi_{2}}d\phi^{\prime}\exp(-\int_{\phi_{*}}^{\phi^{\prime}}Q)\Big)^{-1}.
\end{array}
\end{equation}
$\phi_{*},\phi_{1},\phi_{2}$ are arbitrary but a useful choice is
$\phi_{1}=\phi_{end}$, $\phi_{2}=\infty$ and $\phi_{*}=\phi_{in}$.
So, we calculate the first hitting of a boundary  of an interval
$[\phi_{end},\infty]$ by the process starting from $\phi_{in}$(we
know that $\phi=\infty$ cannot be achieved, so there remains
$\phi_{end}$).  There may be some  problems with integrability in
eq.(36) with some potentials in an infinite interval \cite{
ven1}\cite{ven2} (if there is no thermal noise). Then, $\int
Q\simeq -V^{-1}$ and the integrability may fail if either $V=0$ or
$V$ does not grow fast enough .

 From the formula
(23) we have
\begin{equation}\begin{array}{l}
\nu=f_{I}(\phi_{in})+\int_{0}^{\phi}\partial_{\phi}f_{I}(
\frac{\gamma}{3(aH)^{\frac{3}{2}}}dB(\nu)+\frac{1}{2\pi}
HdW(\nu)).
\end{array}\end{equation}
Taking the square and then the expectation value of eq.(37) we
obtain
\begin{equation}\begin{array}{l}
\langle \nu^{2}\rangle \equiv \langle \nu\rangle^{2}+\langle
(\delta\nu)^{2}\rangle\cr =\langle \nu\rangle^{2}+ \langle
\int_{0}^{\nu}d\nu^{\prime}(\partial_{\phi}f_{I})^{2}(\frac{\gamma^{2}}{9(aH)^{3}}+\frac{H^{2}}{4\pi^{2}})\rangle
\end{array}\end{equation} (an analogous formula holds true for $f_{S}$).
Assume that we find a function $F_{S}$ such that
\begin{equation}\begin{array}{l}
-\partial_{\phi}F_{S}\frac{1}{3H^{2}}V^{\prime} +
\frac{1}{18}\frac{\gamma}{(aH)^{\frac{3}{2}}}\partial_{\phi}\frac{\gamma}{(aH)^{\frac{3}{2}}}\partial_{\phi}F_{S}
+\frac{1}{8\pi^{2}}\partial_{\phi}H\partial_{\phi}HF_{S}\cr=
-(\partial_{\phi}f_{S})^{2}(\frac{\gamma^{2}}{9(aH)^{3}}+\frac{H^{2}}{4\pi^{2}})
\end{array}\end{equation}in the Stratonovitch case and
$F_{I}$
\begin{equation}\begin{array}{l}
-\partial_{\phi}F_{I}\frac{1}{3H^{2}}V^{\prime}+
\frac{1}{18}\Big(\frac{\gamma}{(aH)^{\frac{3}{2}}}\Big)^{2}\partial_{\phi}\partial_{\phi}F_{I}
+\frac{1}{8\pi^{2}}H^{2}\partial_{\phi}\partial_{\phi}F_{I}\cr=
-(\partial_{\phi}f_{I})^{2}(\frac{\gamma^{2}}{9(aH)^{3}}+\frac{H^{2}}{4\pi^{2}})
\end{array}\end{equation}in the Ito case.
Then, calculating $dF$ in the same way as we did for $df$ in
eqs.(22)-(23) using eqs.(38)-(40) and taking the expectation value
we find
\begin{equation}
\langle \nu^{2}\rangle -\langle\nu\rangle^{2}=F(\phi_{in}).
\end{equation}
Let us denote $U=\partial_{\phi}F_{I}$ . Then, in the Ito version
we have the equation
\begin{equation}
\partial_{\phi}U+QU=-R
\end{equation} with $Q$ defined in eq.(28) and
\begin{equation}
R=2(\partial_{\phi}f_{I})^{2}.
\end{equation}
The solution follows the one of eq.(27) as expressed either in
eq.(30) or in eq.(33). In the next section we discuss perturbative
solutions of eqs.(27) and (42).
\section{ e-folds, their fluctuations and the power spectrum}
The general integral formulae in sec.3 do not allow an explicit
calculation of the functions $F$ and $f$ needed for a computation
of the e-folds and their fluctuations. We need a perturbative
approach. We write eq.(27)as an iterative perturbation expansion
in $\frac{1}{Q}$ starting with $u^{(1)}=-\frac{r}{Q}$
\begin{equation} u^{(n+1)}=-\frac{1}{Q}\partial_{\phi}u^{(n)}.
\end{equation} Then, the zeroth order approximation in eq.(27) corresponds to setting
$\frac{1}{Q}\partial_{\phi}u=0$. Hence,\begin{equation}
u^{(1)}=f^{\prime}= -rQ^{-1}=8\pi G V(V^{\prime})^{-1}.
\end{equation} In this approximation we have derived the "classical" formula
(9) for e-folds. We have
\begin{equation}
\partial_{\phi}u^{(1)}=8\pi G(1-\frac{\eta}{2\epsilon})
\end{equation}
with
\begin{equation}
\epsilon=\frac{1}{16\pi G}(\frac{V^{\prime}}{V})^{2}
\end{equation}
\begin{equation}
\eta=\frac{1}{8\pi G}\frac{V^{\prime\prime}}{V}.
\end{equation}Hence, $\partial_{\phi} u^{(1)}\neq 0$ in general.
 We get
$\partial_{\phi} u^{(1)}=0$ for the  exponential potential. For a
power-law potential $\phi^{n}$ we have
$\partial_{\phi}u^{(1)}=\frac{8\pi G}{n}$. $u^{(1)}$ is
independent of $\gamma$ but $u^{(2)}$ (44) depends on $\gamma$ as
$Q$ does. If the thermal noise is absent then the expansion (44)
is an expansion in $G$ (i.e., in the inverse of the Planck mass).
Next, we need an approximation for the solution of eq.(42).
Applying again the expansion (44) in $\frac{1}{Q}$ ( in the lowest
order $\frac{1}{Q}\partial_{\phi}U\simeq 0$) we obtain
\begin{equation}\begin{array}{l}
U^{(1)}=\partial_{\phi}F^{I}\cr=-RQ^{-1}= 6(8\pi
G)^{2}V^{2}(V^{\prime})^{-3}H^{2}(\frac{H^{2}}{8\pi^{2}}+\frac{\gamma^{2}}{18a^{3}H^{3}}).
\end{array}\end{equation} In the next order
\begin{equation}\begin{array}{l}
U^{(2)}=3H^{2}(\frac{H^{2}}{8\pi^{2}}+\frac{\gamma^{2}}{18a^{3}H^{3}})(V^{\prime})^{-1}\partial_{\phi}U^{(1)}.
\end{array}\end{equation}
The power spectrum ${\cal P}\simeq Uu^{-1}$ at $\gamma=0$ derived
from eqs.(45) and (49) coincides with the standard formula
\cite{mukhanov}\cite{starpl}\cite{hawking}
 \cite{guth}\cite{bar}\cite{mu}\cite{sas}\cite{power}
\cite{pot}. It can be obtained from
 the general formulae of sec.3 which involve $\int Q$. These formulae
for e-foldings in the quantum case (cold inflation) have been
discussed by Starobinsky and Vennin \cite{vennini}. If $\gamma=0$
then
\begin{displaymath}
\int_{\psi}^{\phi}Q=-\frac{3}{8G^{2}}\frac{1}{V(\phi)}+\frac{3}{8G^{2}}\frac{1}{V(\psi)}.
\end{displaymath}
In order to calculate the integrals  (30)-(36) they perform the
Taylor expansion of $\frac{1}{V(\psi)}$ around $\phi$
\begin{equation}
\frac{1}{V(\psi)}=\frac{1}{V(\phi)}+\partial_{\phi}\frac{1}{V(\phi)}(\psi-\phi)+...
\end{equation}
Changing variables
\begin{displaymath}
\psi-\phi=G^{2}X \end{displaymath} and expanding the exponential
in $G^{2}$ we derive the perturbation expansion (44).

  Next, let us
consider  the thermal noise using the method of Starobinsky and
 Vennin \cite{vennini}\cite{venninith} (which is equivalent to the expansion (44) in
$\gamma^{2}G^{-\frac{1}{2}}$ and in $G$). Then, in eq.(28)
(without the quantum noise)
\begin{equation}
Q=-\frac{6}{\gamma^{2}}\sqrt{\frac{8\pi
G}{3}}V^{\prime}\sqrt{V}a^{3}
\end{equation}and in eq.(29)
\begin{equation}r=\gamma^{-2} 48\pi G(\frac{3}{8\pi
G})^{\frac{1}{2}}V^{\frac{3}{2}}a^{3}\end{equation} We have by an
integration by parts
\begin{equation}\begin{array}{l}
\int_{\psi}^{\phi}Q=-\frac{4}{\gamma^{2}}\sqrt{\frac{8\pi
G}{3}}\Big(
a^{3}(\phi)V^{\frac{3}{2}}(\phi)\cr-a^{3}(\psi)V^{\frac{3}{2}}(\psi)\Big)
-\frac{32\pi G}{\gamma^{2}}\sqrt{\frac{8\pi
G}{3}}\int_{\psi}^{\phi}\frac{V^{\frac{5}{2}}a^{3}}{V^{\prime}}.
\end{array}\end{equation} Let \begin{equation}
\Omega(\psi)=\frac{4}{\gamma^{2}}\sqrt{\frac{8\pi
G}{3}}a^{3}(\psi)V^{\frac{3}{2}}(\psi).\end{equation} We repeat
the approximation (51) used in \cite{vennini}\cite{venninith} when
applied to eq.(31). We expand in eq.(31) the exponential function
\begin{equation}
\Omega(\psi)=\Omega(\phi)+\partial_{\phi}\Omega(\phi)(\psi-\phi)+...
\end{equation}
Then, in the integral (31) we have (neglecting
$\partial_{\phi}a\simeq G$ and the second term in eq.(54) being of
higher order in $G$)
\begin{equation}
\langle \nu\rangle=\int d\phi r(\phi)\int^{\phi}
d\psi\exp(\partial_{\phi}\Omega(\phi)\psi)\simeq 8\pi G \int
d\phi\frac{V}{V^{\prime}}
\end{equation}
By means of integral formulae of sec.3 as well as with the
perturbation expansion (44) we obtain in the first order the same
formula for e-folds as we could get in the calculation without
noise (showing that the stochastic method of reaching  the
boundary has the correct no noise limit; the stochastic formula
will still  be discussed in Appendix A).The calculation of
$\langle \delta\nu^{2}\rangle$ with the thermal noise on the basis
of eq.(41) involves calculation of the integral
\begin{equation}
F\simeq 2\int^{\phi} d\psi
(f^{\prime}(\psi))^{2}\exp(-\int_{\psi}^{\phi}Q)
\end{equation} with $Q$ of eq.(52).
The Taylor expansion (56) in the integral (58) gives
\begin{equation}
U=F^{\prime}=-\gamma^{2}(8\pi G)^{2}(24\pi
G)^{-\frac{1}{2}}V^{\frac{3}{2}}(V^{\prime})^{-3}a^{-3}.
\end{equation}
The power spectrum can be defined by fluctuations of the e-folds
\begin{equation}
{\cal
P}=\frac{d\langle(\delta\nu)^{2}\rangle}{d\langle\nu\rangle},
\end{equation} where $\langle(\delta\nu)^{2}\rangle$ is defined in eq.(38). We have

\begin{equation}
\frac{d}{d\langle\nu\rangle}=-(f^{\prime})^{-1}\frac{d}{d\phi}.
\end{equation}
Hence
\begin{equation}
{\cal P}=F^{\prime}(f^{\prime})^{-1}\equiv \frac{U}{u}
\end{equation}(evaluated at the horizon crossing $k=aH$ \cite{book}) where $f$ is defined in eqs.(24)-(25)
and $F$ in eqs.(39)-(40) (calculated from eq.(42)). It has been
shown \cite{vennini}\cite{venninith} that the formula (62) in the
expansion (51) (no thermal noise) coincides with the standard one
for the cold inflation
\cite{mukhanov}\cite{bar}\cite{hwang}\cite{mu}\cite{sas}
\cite{power}\cite{book}\cite{pot} because we obtain from
eqs.(41)-(42),(45) and (51)
 \begin{equation}
{\cal P}_{q}=\frac{2G^{2}}{3}8\pi G V^{3}(V^{\prime})^{-2}.
\end{equation}
It is difficult to calculate ${\cal P}$ analytically for general
quantum and thermal noise using the formulae of sec.3. We
calculate the power spectrum with no quantum noise (solely thermal
noise) applying for $F^{\prime}$ the same approximation which we
used in eq.(62) ( for $\langle \nu\rangle$, i.e. for
$f^{\prime}$). Then, from eq.(63)
\begin{equation}
{\cal P}_{th}=8\pi G\gamma^{2}(24\pi
G)^{-\frac{1}{2}}a^{-3}V^{\frac{1}{2}}(V^{\prime})^{-2}.
\end{equation}
From the $\frac{1}{Q}$ expansion using eqs.(45),(49) and (62) we
obtain
\begin{equation}
{\cal P}={\cal P}_{th}+{\cal P}_{q}
\end{equation}(this simple additivity holds true only in the
lowest order of the $\frac{1}{Q}$ expansion as can be seen from
eq.(50)). The spectral index $n_{S}$ can be calculated as a
derivative (61) over $\langle \nu\rangle$ of $\ln{\cal P}$. Then
\begin{equation}
n_{S}-1=- (f^{\prime})^{-1}\frac{d}{d\phi}\ln{\cal
P}=-F^{\prime\prime}(f^{\prime}F^{\prime})^{-1}+f^{\prime\prime}(f^{\prime})^{-2}
\end{equation}
We obtain from eq.(64) (the spectral index for warm inflation is
calculated in \cite{power1}\cite{power2} but under different
assumptions)
\begin{equation}
n_{S}^{th}-1=-3-\epsilon+2\eta.
\end{equation}
 For the quantum stochastic inflation the formula (63) gives
\begin{equation}
n^{q}_{S}-1=-6\epsilon+2\eta
\end{equation}( in agreement with \cite{mu}\cite{sas}\cite{power}\cite{book}\cite{pot}).
On the basis of the $\frac{1}{Q}$ expansion using eqs.(66) and
(65) we obtain for the spectral index of the scalar field in
thermal and quantum noises
\begin{equation}
n_{S}-1=({\cal P}_{q}+{\cal P}_{th})^{-1}\Big({\cal
P}_{q}(n^{q}_{S}-1)+(n_{S}^{th}-1){\cal P}_{th}\Big).
\end{equation}
For a small $\gamma$ from eq.(69) and eqs.(63)-(68)
\begin{displaymath}
n_{S}-1=-6\epsilon+2\eta+\gamma^{2}\sqrt{\frac{3}{32\pi}}a^{-3}(GV)^{-\frac{5}{2}}(-3+5\epsilon).
\end{displaymath}
\section{Summary} The method of a
description of inflation in terms of the fluctuations of  e-folds
(called $\delta N$ method) has been suggested long time ago
\cite{starpl}\cite{starobjettp}\cite{power}, developed recently
and applied to detailed estimates of inflation parameters
\cite{vennini}\cite{ven1}\cite{ven2}
  \cite{stochmulti}\cite{fujita}\cite{wands}\cite{ven}.
In this paper we have extended this formalism to include a thermal
noise. The thermal noise modifies the results on power spectrum.
In principle, the method allows to calculate the inflationary
parameters non-perturbatively for a larger class of potentials .
In the lowest order of a perturbative expansion we have obtained a
formula for the power spectrum which is just a sum of the density
of thermal (proportional to $\gamma^{2}$) and quantum
fluctuations. The spectral index is an average of spectral indices
of thermal and quantum fluctuations with the corresponding power
spectra. The correction to the power spectrum is proportional to
$\gamma^{2}$ (which is small for low temperature in the warm
inflation models0. The spectral index depends on the inflaton
potential. It is measurable in observations \cite{ade}. Its value
can give some information on the inflaton potential as well as on
the friction in inflaton wave equation.

{\bf Acknowledgements} The author thanks the anonymous referees
whose comments contributed to a substantial improvement of the
initial version of this paper.

\section{Appendix A}

It is instructive to compare approximations applied in secs.3-5
with exact solutions (some solutions of stochastic equations with
quantum noise are discussed in \cite{FP1}-\cite{FP4}).

For the exponential potential (see \cite{exp}\cite{mar})
 $V=g\exp(\lambda\phi)$ in the e-folding time
the stochastic equation (6) reads (with $a(\phi)$ derived in
eq.(16)) \begin{equation}\begin{array}{l}
d\phi=-\frac{\lambda}{8\pi G}d\nu+\frac{1}{2\pi}\sqrt{\frac{8\pi
Gg}{3}}\exp(\frac{1}{2}\lambda\phi)\circ
dW(\nu)\cr+\frac{\gamma}{3}(\frac{3}{8\pi
Gg})^{\frac{3}{4}}\exp(\alpha\phi)\circ dB(\nu),
\end{array}\end{equation} where
\begin{displaymath}
\alpha=\frac{12\pi G}{\lambda}-\frac{3}{4}\lambda.
\end{displaymath}
If we express $a$ by $\nu$ in eq.(6) (leading to the Fokker-Planck
equation (14)) then
\begin{equation}\begin{array}{l}
d\phi=-\frac{\lambda}{8\pi G}d\nu+\frac{1}{2\pi}\sqrt{\frac{8\pi
Gg}{3}}\exp(\frac{1}{2}\lambda\phi)\circ
dW(\nu)\cr+\frac{\gamma}{3}(\frac{3}{8\pi
Gg})^{\frac{3}{4}}\exp(-\frac{3}{2}\nu)\exp(-\frac{3}{4}\lambda
\phi)\circ dB(\nu).\end{array}\end{equation} In the Stratonovitch
interpretation these equations can be solved exactly if either the
quantum noise or the thermal noise are absent. No exact solution
exists if the Stratonovitch differential in eqs.(70)-(71) is
replaced by the Ito differential. In such a case in order to
approach the solution we would have to use the relation
\cite{ikeda}(with a certain parameter $\sigma$)
$\exp(\sigma\phi)dW= \exp(\sigma\phi)\circ dW -\frac{1}{2}\sigma
d\phi dW \exp(\sigma\phi)$. The resulting Stratonovitch equation
(which can be treated as an ordinary differential equation) would
not be linear. If quantum noise is absent then we set
\begin{equation}
X=\exp(-\alpha\phi). \end{equation} Then ( in the decomposition
into classical and stochastic
parts)\begin{equation}\begin{array}{l} X\equiv
X_{cl}+X_{st}=\exp(\frac{\alpha\lambda}{8\pi G}\nu)X_{0}
\cr-\frac{\alpha\gamma}{3}(\frac{3}{8\pi Gg})^{\frac{3}{4}}
\int_{0}^{\nu}\exp(\frac{\alpha\lambda}{8\pi
G}(\nu-s))dB(s).\end{array}\end{equation} For eq.(71) we set
\begin{equation}
\tilde{X}=\exp(\frac{3}{4}\lambda\phi).
\end{equation}
Then (no quantum noise)
\begin{equation}\begin{array}{l}
\tilde{X}=\exp(-\frac{3\lambda^{2}}{32\pi
G}\nu)\tilde{X}_{0}\cr+\frac{\lambda\gamma}{4}(\frac{3}{8\pi
Gg})^{\frac{3}{4}}\int_{0}^{\nu}\exp\Big(-\frac{3\lambda^{2}}{32\pi
G}(\nu-s)-\frac{3}{2}s\Big)dB(s) .\end{array}\end{equation} If
$\gamma=0$ in eq.(70) then we set
\begin{equation}
Y=\exp(-\frac{1}{2}\lambda\phi).
\end{equation}
We obtain the solution
\begin{equation}\begin{array}{l}
Y\equiv Y_{cl}+Y_{st}=\exp(\frac{\lambda^{2}}{16\pi
G}\nu)Y_{0}\cr- \frac{\lambda}{4\pi}\sqrt{\frac{8\pi
Gg}{3}}\int_{0}^{\nu}\exp(\frac{\lambda^{2}}{16\pi
G}(\nu-s))dW(s).
\end{array}\end{equation} We can calculate the power spectrum from the
formula for the energy density  fluctuations $\delta\rho$
\begin{equation}
\langle(\frac{\delta\rho}{\rho})^{2}\rangle= \langle \Big(
H(\frac{d\phi}{dt})^{-1}\delta\phi\Big)^{2}\rangle,
\end{equation}
where
\begin{equation}
\delta\phi=\phi-\phi_{cl}.
\end{equation}
For the exponential interaction in the slow-roll approximation
\begin{equation} \Big(\frac{H}{\frac{d\phi}{dt}}\Big)^{2}=(8\pi
G)^{2}\lambda^{-2}.
\end{equation}
Hence, fluctuations of $\frac{\delta\rho}{\rho}$ are proportional
to fluctuations of $\phi$. Fluctuations of $\phi$ in eqs.(73),(75)
and (77) can be calculated  in a power series expansion in the
noise   (see similar calculations in \cite{matacz}) . So, in the
case of the thermal noise \begin{equation} \phi-\phi_{cl}=
\beta\int_{0}^{\nu}ds\exp(-\frac{\lambda\alpha s}{8\pi G})dB(s)
\end{equation}
where \begin{displaymath} \beta= \frac{\gamma}{3}(\frac{3}{8\pi
Gg})^{\frac{3}{4}}\end{displaymath} Hence,
\begin{equation}
\langle (\delta\phi)^{2}\rangle=(\frac{4\beta\pi
G}{\alpha})^{2}(1-\exp(-\frac{\lambda\alpha \nu}{4\pi G})).
\end{equation} In the case of the quantum noise
\begin{equation}
\langle (\delta\phi)^{2}\rangle=\frac{4}{3}\lambda^{-2}(8\pi
G)^{2}(1-\exp(-\frac{\lambda^{2}\nu}{8\pi G}))
\end{equation}(in agreement with eqs.(38)-(39) of \cite{fin2}).
$\nu$ should be taken at the  time $\nu \simeq -\ln k$ expressed
by the wave number $k$ at the Hubble horizon crossing.

 From eq.(9)
\begin{equation}
\nu=\nu_{in}-\frac{8\pi G}{\lambda}\phi.
\end{equation}
Hence, at the lowest order fluctuations of $\phi$ are proportional
to fluctuations of $\nu$ . Now, we can calculate the spectral
index as $-\partial_{\nu}\ln\langle (\delta\phi)^{2}\rangle$ with
the result\begin{equation}
n_{S}^{th}-1=-3+\frac{3\lambda^{2}}{16\pi G}
\end{equation} for thermal noise and
\begin{equation}
n^{q}_{S}-1=-\frac{\lambda^{2}}{8\pi G}\end{equation}for the
Starobinsky (quantum) noise  in agreement with eqs.(67)-(68). In
all cases (73)-(77) we obtain stochastic corrections to the
classical formula (9). As an example from eq.(77)
\begin{displaymath}\begin{array}{l}
\langle \exp(-\frac{1}{2}\lambda\phi)\rangle \simeq\langle
\exp(-\frac{1}{2}\lambda\phi_{cl})\rangle \cr
=\exp(-\frac{1}{2}\lambda\phi_{0})\langle\exp(\frac{\lambda^{2}}{16\pi
G}\nu)\rangle. \end{array}\end{displaymath}Hence, in the
approximation $\langle \exp(f)\rangle\simeq \exp(\langle
f\rangle)$ we obtain the classic formula $\langle \nu\rangle=-8\pi
G\langle \phi\rangle $. From eq. (77) we could obtain further
relations between correlation functions of $\phi$ and $\nu$.

\section{Appendix B}
Let us repeat the derivation of the fluctuation equations (without
the thermal noise) of ref.\cite{fin1}
\cite{fin2}\cite{vennini}with some care concerning the rules of
the stochastic calculus \cite{ikeda}\cite{simon}. We discuss here
also  the difference between the Ito and Stratonovith versions.
The conventional rules of the differential calculus (in
particular, the Leibniz rule) are satisfied in the Stratonovitch
form of stochastic equations. On the other hand correlation
functions are easier to calculate with the Ito integrals (in
particular, an expectation value of the Ito integral is zero). One
can relate both integrals according to the rule
\begin{equation}
f\circ dW=fdW+\frac{1}{2}dfdW,
\end{equation}
where after calculation of $df$ as a function of $W$ we use the
rule $dWdW=d\nu$.

Let
\begin{equation}
\partial_{\nu}\phi_{cl}=-\frac{1}{3}H(\phi_{cl})^{-2}V^{\prime}(\phi_{cl}),
\end{equation}where $H$ is related to $V$ by eq.(8).
Let $\delta=\phi-\phi_{c}$.First, consider the Stratonovitch
version of eq.(6). We write $\phi=\phi_{c}+\delta$, use eq.(88)
and expand eq.(6) till the second order in $\delta$. Then,
integrating, taking the expectation value and using $\langle \int
fdW\rangle=0$ we obtain
\begin{equation}\begin{array}{l}
\langle \delta\rangle^{\prime}= \Big(
(\ln(\frac{H^{\prime}}{H}))^{\prime} -\frac{G}{2
\pi}(H^{\prime}H)^{\prime}\frac{H}{H^{\prime}}\Big)\langle
\delta\rangle\cr+\frac{H}{H^{\prime}}\Big(\frac{1}{2}(\frac{H^{\prime}}{H})^{\prime\prime}-
\frac{G}{4\pi}(HH^{\prime})^{\prime\prime}\Big)\langle
\delta^{2}\rangle-\frac{G}{2\pi}H^{2}.
\end{array}\end{equation}
The equation for fluctuations reads
\begin{equation}\begin{array}{l}\langle
\delta^{2}\rangle^{\prime}=\Big(2(\ln(\frac{H^{\prime}}{H}))^{\prime}
-\frac{G}{\pi}(H^{2})^{\prime\prime}\frac{H}{H^{\prime}}\Big)\langle
\delta^{2}\rangle\cr -\frac{3G}{\pi}H^{2}\langle \delta\rangle
-\frac{G}{\pi}\frac{H^{3}}{H^{\prime}}.
\end{array}\end{equation}
In these equations we have replaced the variable $\nu$ by
$\phi_{cl}$ on the basis of eq.(88). The "prime" denotes a
differentiation with respect to $\phi_{cl}$.

In the Ito interpretation of eq.(6)

\begin{equation}\begin{array}{l}
\langle \delta\rangle^{\prime}=
(\ln(\frac{H^{\prime}}{H}))^{\prime} \langle
\delta\rangle+\frac{1}{2}\frac{H}{H^{\prime}}(\frac{H^{\prime}}{H})^{\prime\prime}\langle
\delta^{2}\rangle,
\end{array}\end{equation}

\begin{equation}\begin{array}{l}\langle
\delta^{2}\rangle^{\prime}=\Big(2(\ln(\frac{H^{\prime}}{H}))^{\prime}
-\frac{G}{2\pi}(H^{2})^{\prime\prime}\frac{H}{H^{\prime}}\Big)\langle
\delta^{2}\rangle\cr -\frac{2G}{\pi}H^{2}\langle \delta\rangle
-\frac{G}{\pi}\frac{H^{3}}{H^{\prime}}.
\end{array}\end{equation}
Eq.(92) is different from eq.(38) of \cite{fin1} and eq.(A.22) of
\cite{vennini} as the terms $GH^{2}\langle\delta\rangle$ and
$G(H^{2})^{\prime\prime}$ are absent there. We can solve both
Stratonovitch (89)-(90) and Ito (91)-(92) equations in a
perturbation expansion in $m_{pl}^{-2}=8\pi G$. We expand the
solution of eqs.(91)-(92) around the one of
\cite{fin1}\cite{vennini} (with $\langle\delta\rangle=0$ and no
extra terms) then we obtain that in such an expansion
$\langle\delta\rangle\simeq G^{2}$. Hence, the term
$GH^{2}\langle\delta\rangle$ in eq.(92) will be of order $G^{4}$.
As $\langle \delta^{2}\rangle\simeq G^{2}$ another extra term in
eq.(92)$G(H^{2})^{\prime\prime}\langle \delta^{2}\rangle\simeq
G^{4}$. Hence, in comparison with refs.\cite{fin1} \cite{vennini}
the extra terms are of higher order in $G$ (at this order, our
starting point, the equations for stochastic imflation would also
need a modification). The same argument applies to the
Stratonovitch eq.(90). It is different from equations of
refs.\cite{fin1} \cite{vennini} by terms of order $G^{4}$. For
this reason till the order $G^{4}$ we have the same conclusions
concerning the solution of eqs.(90) or (92) (these equations
determine the fluctuations and power spectra).

We can repeat the calculations of fluctuations for the thermal
noise rederiving the formula (64). Then, at the lowest order in
$\gamma^{2}$ we get the additivity of fluctuations (65) and as a
consequence the formula (69) for the spectral index.

 Let us consider  the particular case of the Stratonovitch stochastic equation
(6) for $V=\frac{1}{2}m^{2}\phi^{2}$ with no thermal noise
\begin{equation} d\phi=-\frac{d\nu}{4\pi G\phi}
+q\phi\circ dW(\nu),
\end{equation}
where $ q=\sqrt{\frac{Gm^{2}}{3\pi}}$. The Ito version of eq.(6)
is
\begin{equation} d\phi=-\frac{d\nu}{4\pi G\phi}
+q\phi dW(\nu)=-\frac{d\nu}{4\pi G\phi}-\frac{1}{2}q^{2}\phi d\nu
+q\phi\circ dW(\nu)
\end{equation}
Eq.(93)  has the solution \begin{equation}
\phi_{\nu}^{2}=\phi_{0}^{2}\exp(2qW(\nu))-\frac{1}{2\pi
G}\int_{0}^{\nu}ds\exp\Big(2qW(\nu)-2qW(s)\Big)
\end{equation}where $\phi_{0}$ is the initial value.
The solution of eq.(94) reads
\begin{equation}\begin{array}{l}
\phi_{\nu}^{2}=\phi_{0}^{2}\exp(-q^{2}\nu+2qW(\nu))\cr-\frac{1}{2\pi
G}\int_{0}^{\nu}ds\exp\Big(-q^{2}(\nu-s)+2qW(\nu)-2qW(s)\Big).
\end{array}\end{equation}

We can express correlation functions of $\phi^{2}$ in terms of
correlations of $\nu$. In particular, for the Stratonovitch
version
\begin{equation}
 \langle\phi_{\nu}^{2}\rangle
=(\phi_{0}^{2}- q^{-2}\frac{1}{4\pi G})\langle
\exp(2q^{2}\nu)\rangle + q^{-2}\frac{1}{4\pi G}
\end{equation} and for the Ito version
\begin{equation}
 \langle\phi_{\nu}^{2}\rangle
=(\phi_{0}^{2}- q^{-2}\frac{1}{2\pi G})\langle
\exp(q^{2}\nu)\rangle + q^{-2}\frac{1}{2\pi G}.
\end{equation}
It can be seen that from the requirement of positivity of
$\phi_{\nu}^{2}$ we get  some bounds on the initial values
$\phi_{0}$ and expectation values of $\nu$. In particular, the
relation (15) $\langle \phi_{\nu}^{2}\rangle=const -\frac{1}{2\pi
G}\langle \nu\rangle$ holds true only in the lowest order in $q$.
The fluctuation equations for the model $
\frac{1}{2}m^{2}\phi^{2}$ discussed in \cite{vennini} (eq.(A.33))
follow from eqs.(89) or (92). The exact solution (95)-(96) does
not tell us more than eqs.(89)-(92) (with $H\simeq \phi$)
concerning quadratic fluctuations. However, using the solutions
(95)-(96) we could get explicitly the higher order fluctuations of
$\phi^{2}$.

\end{document}